\documentclass[preprint,12pt]{elsarticle}

\usepackage{amsthm,amscd}
 \usepackage{amsmath}
\usepackage{amssymb}
\usepackage{lipsum}

\newcommand{\rme}{\mathrm{e\,}}

\def\p{\partial}

\journal{Annals of Physics}

\begin{document}

\begin{frontmatter}



\title{Propagators of singular anharmonic oscillators with quasi-equidistant spectra}


\author[first]{Andrey M. Pupasov-Maksimov}
\affiliation[first]{organization={Federal University of Juiz de Fora},
            addressline={ICE, Dept. Mat.}, 
            city={Juiz de Fora},
            state={MG},
            country={Brasil}}

\author[second]{Marcelo Silva Oliveira}
\affiliation[second]{organization={Bahia State Department of Education},
            city={Salvador},
            state={BA},
            country={Brasil}}

\begin{abstract}
Darboux transformations of the singular harmonic oscillator are considered. 
Analytical expressions for the propagators are obtained, using 
the image method applied to formal singular propagators. 
Two-well and three-well families of potentials and the corresponding propagators are presented. 
Axially symmetric magnetic field configurations corresponding to these potentials have been identified.
\end{abstract}



\begin{keyword}
Propagators \sep Darboux transformations \sep Singular harmonic oscillator \sep Image method \sep Aharonov–Bohm flux \sep axial magnetic fields



\end{keyword}

\end{frontmatter}




\section{Introduction}

This work is devoted to the analytical construction of propagators for the one-dimensional time-dependent Schr{\" o}dinger equation with rationally extended potentials. The propagator, a fundamental solution of the Cauchy problem, encodes the full dynamics of a quantum system and determines the transition amplitude between quantum states. It can be obtained either via an expansion over stationary states or, alternatively, through the path-integral formulation introduced by Feynman \cite{feynman1948, feynman2010}, which plays a central role in quantum mechanics. An extensive list of exact propagators as well as exact and approximate techniques for their calculations can be found in \cite{grosche1998handbook}.

Our focus lies on quantum systems obtained via Darboux transformations of the harmonic oscillator. These transformations generate exactly solvable potentials with modified spectral properties. Although most of the spectrum remains equidistant, at low energies 
eigenvalues may be shifted.
At the same time, resulting potentials manifest multi-well structures. When only rational extensions \cite{carinena2016ground,carinena2017abc} are considered, the transition energies between levels are multiples of the energy quantum. Surprisingly, the associated propagators admit closed-form expressions in elementary functions only \cite{pupasov2015propagators}. In the present work we extend our previous findings by including the case of singular 
oscillator and its Darboux transformations. 

A Darboux transformation of a regular potential that results in a transformed potential with a singularity, in general, leads to a quantum system with a completely different spectrum.
Moreover, it is necessary to specify domain of definition 
for the transformed Hamiltonian. Again, in general there can be an ambiguity. However, in the case of the harmonic oscillator, Darboux transformations leading to a single singularity at the origin, result in the potentials and hamiltonians which permit a natural interpretation in terms of radial Schr{\" o}dinger equation for a fixed partial wave. 


In the final part of this work, we provide a physical embedding for these singular models by demonstrating that the potentials with inverse-square singularities naturally emerge from the radial reduction of a three-dimensional Schr{\" o}dinger equation in the presence of an axially symmetric magnetic field and an Aharonov--Bohm flux. This connection  highlights the relevance of our constructions for realistic quantum systems with cylindrical symmetry.


\subsection{Notations}
To avoid nested subscripts, we will adopt a new notation for elements of a sequence. We will denote the $i$-th element of a sequence $\sigma$ as $\sigma [[i]]$ and the last element as $\sigma [[-1]]$.
Sequences (or sets) of functions \\
$\{\psi_{n_1}(x),\psi_{n_2}(x),\ldots,\psi_{n_{N}}(x)\}$
will be written compactly as $\psi_{\{n_1,n_2,\ldots,n_N\}}=\psi_{\sigma}(x)$.
With this notation Wronskians involved in Darboux transformations read 
\begin{equation}
{\rm Wr}[\psi_{\sigma}(x),x]=
\left|\begin{array}{cccc}
\psi_{\sigma[[1]]}(x) & \psi_{\sigma[[2]]}(x) & \ldots &\psi_{\sigma[[-1]]}(x)\\
\psi'_{\sigma[[1]]}(x) & \psi'_{\sigma[[2]]}(x) & \ldots &\psi'_{\sigma[[-1]]}(x)\\
\ldots&\ldots&\ldots&\ldots\\
\psi_{\sigma[[1]]}^{(|\sigma|-1)}(x) & \psi_{\sigma[[2]]}^{(|\sigma|-1)}(x) & \ldots &\psi_{\sigma[[-1]]}^{(|\sigma|-1)}(x)\\
\end{array}\right|.
\end{equation}
We use $\cup$ to 
define concatenation 
$$\psi_{\sigma}(x)\cup \{f(x)\}=\{\psi_{\sigma[[1]]}(x),\ldots , \psi_{\sigma[[-1]]}(x),f(x)\}\,.$$
A sequence $\sigma$ with an $n$th element skipped reads $\sigma\setminus \{\sigma[[n]]\}$.

\section{Singular Harmonic oscillator and its rational extensions}
\subsection{Darboux transformation}
Rational extensions of the Harmonic oscillator 
\begin{eqnarray}\label{def:Harmonic-oscillator-Hamiltonian-dimensionless-form}
H_{\rm osc}=-\partial^2_{x}+\frac{x^2}{4}\,,
\end{eqnarray}
are defined 
with the aid of an $N$th order Darboux transformation \cite{bagrov1995darboux} with eigen-functions of the Harmonic oscillator as 
seed solutions.
The oscillator is perturbed by a potential written in terms of the second logarithmic derivative of the Wronskian of the seed solutions
\begin{eqnarray}\label{def:rational-extensions-of-Harmonic-oscillator}
H^{\sigma}=-\p^2_{x}+\frac{x^2}{4}-2\p^2_{x}(\ln {\rm Wr}[\psi_\sigma(x),x])\,.
\end{eqnarray}
Here a string of natural numbers $\sigma=\{n_1,n_2\ldots, n_N\}$ denotes the seed solutions used for the Darboux transformation. 
Note that eigen-functions of \eqref{def:Harmonic-oscillator-Hamiltonian-dimensionless-form} 
are expressed in terms of probabilistic Hermite polynomials
\begin{equation}\label{def:oscillator-eigen-functions}
\psi_n(x)=p_n \mathrm{He}_n(x)\rme^{-\frac{x^2}{4}}\,, \qquad p_n=\left(n!\sqrt{2\pi} \right)^{-\frac{1}{2}}\,.
\end{equation}

Hamiltonians $H_{\rm osc}$ and $H^{\sigma}$ can {\it formally} be embedded into a polynomial SUSY algebra 
\cite{andrianov2012nonlinear}
\begin{eqnarray}\label{ident:intertwinning-relation}
LH_{\rm osc} = H^\sigma L\,,\quad L^+L = \prod\limits_{j=1}^{2M}(H_{\rm osc}-\sigma[[j]])\,,\quad LL^+ = \prod\limits_{j=1}^{2M}(H^{\sigma}-\sigma[[j]])\,,
\end{eqnarray}
where $L$ is a differential operator 
 of $2M$-th order \cite{bagrov1995darboux}
\begin{eqnarray}\label{def:intertwinning-operator}
L f(x)=\frac{{\rm Wr}[\psi_{\sigma}(x)\cup \{f(x)\},x]}{{\rm Wr}[\psi_\sigma(x),x]}\,.
\end{eqnarray}

In the case when ${\rm Wr}[\psi_\sigma(x),x](x=0)=0$  operators $H_{\rm osc}$ and $H^{\sigma}$ have different domains.
The corresponding Hilbert spaces $\mathcal{L}^2(\mathbb{R})$ and $\mathcal{L}^2(\mathbb{R}_{\geq 0})$ are also different. Therefore, 
some additional analysis is required.

Gaussian exponent ${\rm exp}(-x^2/4)$ of oscillator eigen-functions \eqref{def:oscillator-eigen-functions}
vanishes in \eqref{def:rational-extensions-of-Harmonic-oscillator} 
and transformed potential is expressed by the Wronskian of probabilistic Hermite
polynomials only
\begin{equation}
V^\sigma(x) = \frac{x^2}{4}-2\p^2_{x}(\ln {\rm Wr}[\mathrm{He}_\sigma(x),x])+|\sigma|\,.
\label{eq:Vsigma}
\end{equation}

Wronskians of normalized probabilistic Hermite polynomials $h_n(x)$ 
\begin{equation}\label{def:norm-Hermite-polynom}
h_n(x)=p_n \mathrm{He}_n(x)\,,
\end{equation}
define some exceptional polynomials 
\begin{equation}\label{def:norm-xHermite-polynom}
h_n^{\sigma}(x)=N_n {\rm Wr}[h_{\sigma}\cup h_n,x]\,,
\end{equation}
and normalization factor $N_n$ is given in the next section.

Finally, we define compact notations which can be used when $\sigma$ is fixed
\begin{eqnarray}\label{def:Wronskian-sigma-compact}
W(x)&=&{\rm Wr}[\psi_{\sigma}(x),x]\,,\qquad \hat W(x)={\rm Wr}[\mathrm{He}_{\sigma}(x),x]\,,
\\
\label{def:Wronskian-sigma-n-compact}
W_n(x)&=&{\rm Wr}[\psi_{\sigma\setminus \{\sigma[[n]]\}}(x),x]\,,\qquad 
\hat W_n(x)={\rm Wr}[\mathrm{He}_{\sigma\setminus \{\sigma[[n]]\}}(x),x]\,,
\\
\label{def:hat-L-compact}
\hat L f&=&\frac{{\rm Wr}[\mathrm{He}_{\sigma}(x)\cup\{f\},x]}{{\rm Wr}[\mathrm{He}_\sigma(x),x]}\,.
\end{eqnarray}

\subsection{Singular Harmonic oscillator and its rational extensions \label{sec:sho-def}}
When the so-called gap sequence 
$\sigma$, 
\begin{eqnarray}\label{def:adler-sequence}
\sigma=\{k_1,k_1+1,\ldots,k_{M},k_{M}+1\}\,,\qquad |\sigma|=2M\,,
\end{eqnarray}
is a strictly
increasing sequence of natural numbers grouped in consecutive pairs\footnote{The case of $\sigma$ started with a sequence $\{0,1,2,\ldots,k_0,k_1,k_1+1,\ldots,k_{M},k_{M}+1\}$ will define (up to a constant) the same potential as a sequence $\{k_1-k_0,k_1-k_0+1,\ldots,k_{M}-k_0,k_{M}-k_0+1\}$    }, or a Krein-Adler sequence \cite{adler1994modification}, the rationally extended potential \eqref{def:rational-extensions-of-Harmonic-oscillator} 
is regular for all $x\in \mathbb{R}$. Here we will relax this requirement considering potentials 
with a singularity at the origin, and regular for $x>0$.
Therefore, the Wronskian may vanish at the origin, ${\rm Wr}[\psi_{\sigma}(x),x]|_{x=0}=0$.

\begin{figure}
    \centering
    \includegraphics[width=0.85\linewidth]{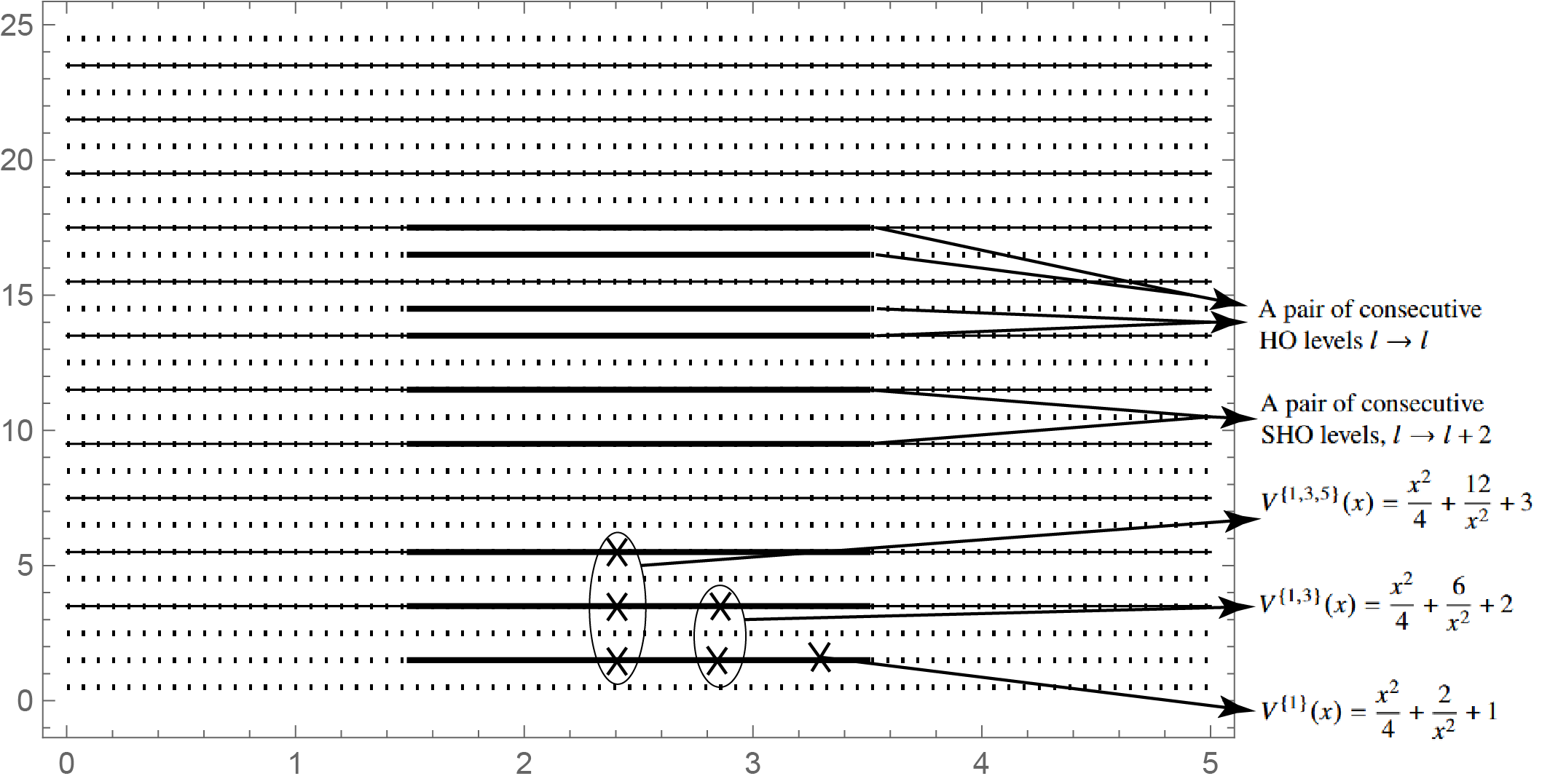}
    \caption{Level diagram of Darboux transformations from the Harmonic oscillator (HO) to the sigular Harmonic oscillator (SHO) and its rational quasi-isospectral
    deformations. Dotted lines show HO eigenvalues, thin black lines show eigenvalues of $H^{\{1\}}$. A possible gap sequence of an arbitrary Darboux transformation 
    are shown by thick black segments.}
    \label{fig:dt-level-diagramm}
\end{figure}

In terms of the initial HO spectrum, the gap sequence 
contains 3 non-intersecting structures
\begin{eqnarray}\label{def:singular-adler-sequence}
\sigma=\sigma_{sing} \cup \sigma_{sing-Adler} \cup \sigma_{Adler} \,,\qquad \qquad\qquad\qquad\qquad\quad\quad \\ 
\sigma_{sing}\cap \sigma_{sing-Adler}=\sigma_{sing-Adler}\cap \sigma_{Adler}=\sigma_{sing}\cap \sigma_{Adler}=\emptyset \,.
\end{eqnarray}
which are shown in 
Figure \ref{fig:dt-level-diagramm}.

The gap sequence $\sigma_{sing}$ contains consecutive odd levels starting from 1.
Resulting Darboux transformations generate singular Harmonic oscillators with different angular momenta
\begin{equation}\label{def:singular-HO}
V^{\{1,3,\ldots,2s+1\}}(x)=\frac{x^2}{4}+\frac{l_s(l_s+1)}{x^2}+l_s\,,\quad l_s=s+1\,.
\end{equation}
Let us consider $V^{\{1,3,\ldots,2s+1\}}(x)$ as a starting potential for the Darboux transformations.
Since ${\rm spec}\, H^{\{1,3,\ldots,2s+1\}}=\{(2s+1+2n)_{n\in \mathbf{N^{*}}}\}$, a Krein-Adler
sequences $\sigma_{sing-Adler}$, chosen from such a spectrum will generate regular potentials for $x>0$. 
Finally, we can start not from the HO but from its 
rational extensions defined by pairs of consecutive levels $\sigma_{Adler}$. 
Then, we can extend this transformation chain by Darboux transformations 
with $\sigma_{sing}$ and $\sigma_{sing-Adler}$ to generate a singular potential.
Thus, an acceptable gap sequence \eqref{def:singular-adler-sequence} is formed from the following  parts
\begin{eqnarray}
    \sigma_{sing} & = & \{1,3,\ldots,2 s+1\}\,, \\
    \sigma_{sing-Adler} & = &\{2 k_1+1,2 k_1+3, \ldots ,2 k_{s}+1,2 k_{s}+3\}\,,\\
    \sigma_{Adler} & = &\{2 m_{1},2 m_{1}+1,\ldots ,2 m_{a},2 m_{a}+1\}\,.
\end{eqnarray}

The transformed potential  $V^{\sigma}(x)$ 
acquires a centrifugal singularity at the origin
\begin{equation}
\lim_{x\to 0}V^{\sigma}(x)x^2=l_N(l_N+1)\,, l_N\in \mathrm{Z}_+\,.
\end{equation}
Integer $l_N$ is defined by the sequence of seed solutions
\begin{equation}\label{def:l-quantum-number}
l_N(\sigma)=|\sigma_{sing}|+|\sigma_{sing-Adler}|\,.
\end{equation}
For example, $\sigma=\{1\}$ gives $l_N=1$, $\sigma=\{1,3\}$ gives $l_N=2$, and
$\sigma=\{1,3,4\}$ gives $l_N=1$. In other words $l_N=\text{number of odd levels-number of even levels}$, 
taking into account that singular deformations of the harmonic oscillator 
allow even levels only in pairs $\{\ldots, 2k, 2k+1, \ldots\}$ or $\{\ldots, 2k-1, 2k, \ldots\}$ with odd levels.

Operator $L$ maps only the odd oscillator eigen-functions \eqref{def:oscillator-eigen-functions}
to the eigen-functions of the rationally extended singular oscillator 
\begin{equation}\label{def:Rext-oscillator-eigen-functions}
\psi_{2k+1}^\sigma(x)=\sqrt{2}N_{2k+1}L\psi_{2k+1}(x)\,,
\end{equation}
where  $\psi_{2k+1}^\sigma(x)\in \mathcal{L}^2(\mathbb{R}_{\geq 0})$ while  $\psi_{2k+1}\in \mathcal{L}^2(\mathbb{R})$.
Functions $L\psi_{2k}(x)\notin \mathcal{L}^2(\mathbb{R}_{\geq 0})$ since $L\psi_{2k}(x\to 0)\to 1/x$.
As a result, we should consider restriction of $L$ to the linear span ${\rm span} \{\psi_{2k+1}(x), k\in \mathbb{N}_0  \}$. 

A normalization factor calculated from \eqref{ident:intertwinning-relation}
\begin{eqnarray}\label{def:normalization-factor}
N_{n}=\left\lbrace 
\begin{array}{cc}
\left( \prod\limits_{j=1}^{|\sigma|}(n-\sigma[[j]])\right) ^{-\frac{1}{2}}\,, & n\notin\sigma\,,\\
0\,, & n\in \sigma\,,
\end{array}
\right. 
\end{eqnarray}
is real for the odd oscillator eigenvalues $n=2k+1$. For the even eigenvalues $N_{2k}$ becomes complex, but these eigenvalues of HO 
are absent in the spectrum of SHO.
Spectrum of the singular rational extension contains only odd levels of the oscillator spectrum except 
(odd) levels from the gap sequence
${\rm spec} H^\sigma=\{3+1/2,\ldots,2k+1+1/2,\ldots\}\setminus \{\text {odd levels from the gap sequence}\}$. 

\begin{figure}
    \centering
    \includegraphics[width=0.85\linewidth]{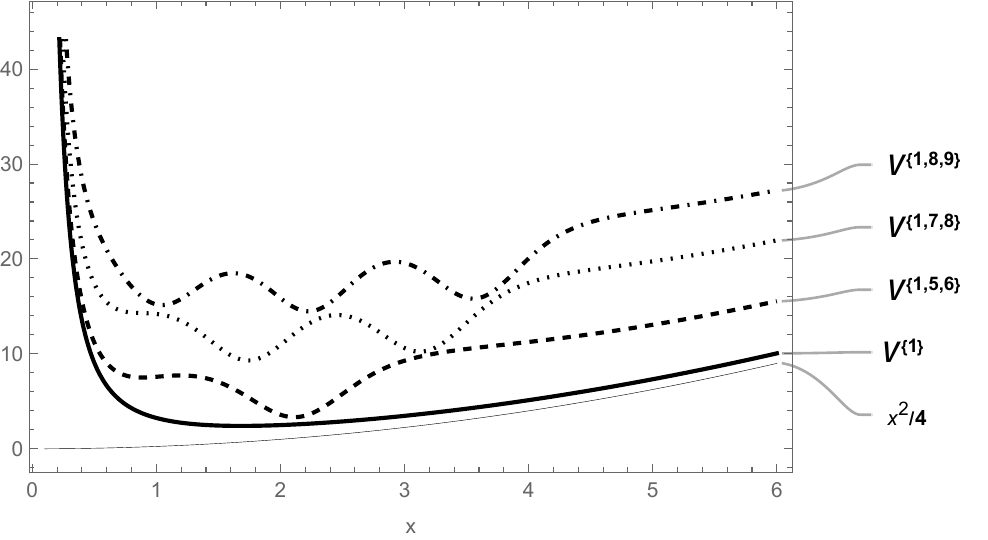}
    \caption{Examples of potentials }
    \label{fig:SHO-1-2k-2kp1}
\end{figure}

Singular Harmonic oscillator \eqref{def:singular-HO} belongs to the class of shape-invariant potentials \cite{gendenshtein1983derivation,grandati2011solvable}.
They have a unique classical equilibrium which is shifted to larger $x$ with increasing $s$. 
A general choice of $\sigma$ produces potentials without shape invariance.  
In figure \ref{fig:SHO-1-2k-2kp1} we plot examples of singular oscillators with multiple potential wells. 

The simplest two-well potential (shown in figure \ref{fig:V-B-1-6-7})
\begin{equation}
\begin{aligned}
 V^{\{1,6,7\}}(x)= \frac{x^2}{4}+\frac{2}{x^2}+\frac{4 \left(5 x^8-26 x^6+8 x^4-1798 x^2-16273\right)}{x^{10}-13 x^8+66 x^6-42 x^4+105 x^2+315}\\
 -\frac{32 \left(12017 x^8-137040 x^6+101682 x^4-295680
   x^2-636615\right)}{\left(x^{10}-13 x^8+66 x^6-42 x^4+105 x^2+315\right)^2}+3
   \label{eq:V-1-6-7}
\end{aligned}
\end{equation}
has the ground $E_0=3+1/2$ and first excited $E_1=5+1/2$ states separated from the equidistant part of the spectrum $E_n=5+1/2+2n$, $n=2,3,\ldots$ by the gap $\Delta E_g=4$. 

The simplest three-well potential is given by
\begin{equation}
\begin{aligned}
    &V^{\{1,8,9\}}(x) = \frac{x^2}{4} + \frac{2}{x^2} + 3 \\
    &+ \frac{4 (-7200549 - 746064 x^2 - 46791 x^4 - 6828 x^6 + 909 x^8 - 
   132 x^{10} + 7 x^{12})}{14175 + 4725 x^2 - 2835 x^4 + 6075 x^6 - 
 2355 x^8 + 423 x^{10} - 33 x^{12} + x^{14}}\\
 &-\frac{5184 (-78704325 - 34615350 x^2 + 13526415 x^4 - 33010860 x^6
   )}{(14175 + 4725 x^2 - 2835 x^4 + 6075 x^6 - 2355 x^8 + 423 x^{10} - 
  33 x^{12} + x^{14})^2}\\
  &-\frac{5184 (9811005 x^8 - 1255862 x^{10} + 31305 x^{12})}{(14175 + 4725 x^2 - 2835 x^4 + 6075 x^6 - 2355 x^8 + 423 x^{10} - 
  33 x^{12} + x^{14})^2}
\end{aligned}
\end{equation}
The ground state, the first excited state, and the second excited state have energies 
$E_0 = 3+1/2$, $E_1 = 5+1/2$, and $E_2 = 7+1/2$, respectively.
These three lowest levels are separated from the quasi-equidistant part of the spectrum 
 $E_n = 5+1/2 + 2n, \quad n = 3, 4, 5, \ldots$
by an energy gap of $\Delta E_g = 4$. In this case, three odd levels of the initial oscillator survive below the quasi-equidistant region, 
resulting in a three-well structure of the potential. 

In general, the simplest \(N_{\mathrm{w}}\)-well potential, with \(N_{\mathrm{w}}\) denoting the number of wells, 
is generated by the sequence
\[
\sigma = \{1,\,2m,\,2m+1\}, \qquad m = 2,3,4, \dots.
\]
The potentials obtained for 
\(\sigma = \{1,4,5\}\), \(\sigma = \{1,6,7\}\), and \(\sigma = \{1,8,9\}\)
correspond to one-, two-, and three-well structures, respectively.
This behavior extends naturally to the general case defined by the sequence above. For this case, the number of wells increases linearly with \(m\) as
$$
N_{\mathrm{w}} = m - 1.
$$
Indeed, the potential $V^{\{2m,\,2m+1\}}(x)$ has $2m$ minima \cite{bagrov1995darboux}, symmetric with respect to $x=0$. Hence, there are 
$m$ minima for $x>0$. 
An additional Darboux transformation  with $\psi_1^{\{2m,\,2m+1\}}(x)$ as a seed solution reduces number of minima by $1$.

This is also in agreement with the spectral structure: 
for a given sequence \(\sigma = \{1,2m,2m+1\}\), there are \(m-1\) eigen-states of the deformed oscillator below 
the equidistant part of the spectrum.


Higher-order rational extensions provide more possibilities for the positions of energy levels below the equidistant
region. 
The sequence $\sigma = \{1,3,6,7,10,11\}$ generates a double-well potential with 
the energy splitting between the ground and the first excited state $\Delta E_{1,0} = 4$ and the gap $\Delta E_{2,1} = 4$ . 
The sequence $\sigma = \{1,3,6,7,10,11,14,15\}$ yields a triple-well configuration
characterized by $\Delta E_{1,0} = \Delta E_{2,1} = \Delta E_{3,2} = 4$.
Such multi-gap configurations were analyzed by \cite{pupasov2016analytical} for the regular in $\mathbb{R}$ potentials.


\section{Darboux transformations for propagators \label{sec:dt-prorpagators} }
In quantum mechanics, the Green function, or propagator, of the Schr\"{o}dinger equation is represented as the kernel of the evolution operator:
\begin{equation}
    K(x,y;t)=\langle x| \hat{U}(t)|y \rangle
\end{equation}
It satisfies a differential
equation with the Dirac delta function as the initial condition
\begin{equation}\label{K00}
[i\p_t-\hat H(x)]K(x,y;t)=0\qquad
K(x,y;0)=\delta(x-y)\,.
\end{equation}
Using the basis of the eigen-functions of the Hamiltonian one can expand the propagator in the following form
\begin{equation}\label{decomp}
K(x,y;t)=
\sum_{n=0}^{\infty}\psi_n(x)\psi_n(y)\mathrm{e}^{-iE_nt}\,.
\end{equation}

We consider the time-dependent Schr\"{o}dinger equation for a rationally extended HO constructed via Darboux transformations.
Let us first recall what happens when $\sigma$ is a Krein-Adler sequence.
The potential is regular for all $x\in \mathbb{R}$
and the corresponding propagator reads as follows \cite{pupasov2015propagators}
\begin{equation}\label{def:rationally-extended-propagator}
K^\sigma(x, y; t) = K_{\text{osc}}(x, y; t) \,
\frac{\displaystyle \sum_{k=0}^{\sigma[[-1]]+1} Q_k^\sigma(x, y) e^{-i k t}}
{\displaystyle \sum_{k=0}^{\sigma[[-1]]+1} Q_k^\sigma(x, y)},
\end{equation}
where $K_{\text{osc}}(x, y; t)$ is the propagator of the harmonic oscillator:
\begin{equation}
K_{\text{osc}}(x, y; t) = 
\frac{1}{\sqrt{4\pi i \sin t}}
\exp\!\left[\frac{i}{4\sin t}\left((x^2+y^2)\cos t - 2xy \right)\right].
\label{eq:Kosc}
\end{equation}

Finite sequence of the polynomials $Q_k^\sigma(x, y)$ should be computed recursively from the relation
\begin{equation}
\begin{aligned}
    Q_k^\sigma(x, y) = \frac{1}{h_0(x)h_0(y)} 
\left(h_k^\sigma(x) h_k^\sigma(y) - \sum_{j=1}^{k} Q_{k-j}^\sigma(x, y) h_j(x) h_j(y)\right),\\
\qquad 0 \le k \le \sigma[[-1]]\ + 1,
\label{eq:Qk}
\end{aligned}
\end{equation}
where $h_n(x)$ are the normalized Hermite polynomials of the harmonic oscillator, 
and $h_n^\sigma(x)$ are so-called (exceptional) x-Hermite polynomials. 
Sum of $Q$-polynomials is equal to the product of Wronskians   
$$
{\displaystyle \sum_{k=0}^{\sigma[[-1]]+1} Q_k^\sigma(x, y)}=\hat{W}(x)\hat{W}(y)\,.
$$

Expression \eqref{def:rationally-extended-propagator} is obtained from the general transformation formula for propagatos
related by Darboux transformation
\cite{pupasov2007exact}.
If two Hamiltonians, $H_0$ and $H_N$, are related by an $N$-th order 
Darboux transformation that removes $N$ levels from the spectrum of 
$H_0$, then the corresponding propagators, $K_0$ and $K_N$, are related as follows\footnote{In our case $b=+\infty$.}
\cite{pupasov2007exact},
\begin{eqnarray}\label{theorem:transformation-GF-general}
K_N(x,y;t)
     & = & L_{x}
  \sum_{n=1}^{N} (-1)^{n}
  \frac{W_{n}(y)}{W(y)}
  \int_{y}^b K_0(x,z;t)u_n(z)dz\,.
\end{eqnarray}
Special properties of Hermite polynomials allow to calculate integrals in \eqref{theorem:transformation-GF-general} explicitly.

Now, if we pass to the radial equation for the SHO, boundary conditions for the wave-functions are also in order for the 
propagator
\begin{equation}\label{def:propagator-SHO-bc}
   K(0,y;t)=K(x,0;t)=0\,.
\end{equation}
Therefore, a direct application of \eqref{def:rationally-extended-propagator}  to calculate $K^\sigma(x, y; t)$ fails. 
Since $W(x)$ has a zero at the origin, transformed propagator \eqref{def:rationally-extended-propagator} 
has singularities when $x,y\to 0$. Nevertheless, $K^\sigma(x, y; t)$ be a formal solution to 
the time-dependent Schr\"{o}dinger equation. Indeed, it is constructed as a sum of Darboux transformation of solutions
which are formal solutions of transformed equation. In the next section we provide a regularization to the  $K^\sigma(x, y; t)$
thus obtaining expressions for the propagator.

\subsection{Image Method for Rational Extensions}
\label{sec:image_method}
As it was noted in \cite{jayannavar1993propagators} in respect to the shape-invariant potentials \cite{das1990propagators}, to obtain the correct propagator for the singular potential
$1/x^2$, we must use the initial propagator on the
half-line. This propagator can be obtained by the image method from the propagator in the full line, due to the symmetry 
of the oscillator Hamiltonian with respect to the reflection $x\to -x$
\begin{equation}
    K_0(x,y;t)=K_{\text{osc}}(x, y; t)-K_{\text{osc}}(-x,y; t)\,.
\end{equation}
Substituting in \eqref{theorem:transformation-GF-general} we get
\begin{eqnarray}\label{theorem:transformation-GF-singular}
K^\sigma(x,y;t)
      =  \\
      L_{x} 
  \sum_{n=1}^{N} (-1)^{n}
  \frac{W_{n}(y)}{W(y)}
  \int_{y}^\infty \left(K_{\text{osc}}(x,z; t)-K_{\text{osc}}(-x,z; t)\right)\psi_{\sigma[[n]]}(z)dz\,.
\end{eqnarray}

Introducing a function 
\[
K_s^{\sigma}(x,y;t)=
L_x\sum_{n=1}^{|\sigma|} (-1)^{n}
  \frac{W_{n}(y)}{W(y)}
  \int_{y}^\infty K_{\text{osc}}(x,z; t) \psi_{\sigma[[n]]}(z)dz\,,
\]
we note that $L_x(x\to-x)=(-1)^{l_N}L_x(x)$, therefore
\[ L_{x} 
  \sum_{n=1}^{N} (-1)^{n}
  \frac{W_{n}(y)}{W(y)}
  \int_{y}^\infty K_{\text{osc}}(-x,z; t)\psi_{\sigma[[n]]}(z)dz=(-1)^{l_N}K_s^{\sigma}(-x,y;t)
\]

Function $K_s^{\sigma}(x,y;t)$ is a formal solution of the time-dependent Schr\"{o}dinger equation with the 
Hamiltonian $H^\sigma$. Since integrals which appear in this function are convergent and has the same structure 
as in the case of rational deformations of the Harmonic oscillator, we can apply the same method as in \cite{pupasov2015propagators}
and express 
\begin{equation}\label{def:sing-extended-propagator}
K_s^\sigma(x, y; t) = K_{\text{osc}}(x, y; t) \,
\frac{\displaystyle \sum_{k=0}^{\sigma[[[-1]]+1} Q_k^\sigma(x, y) e^{-i k t}}
{\displaystyle \sum_{k=0}^{\sigma[[[-1]]+1} Q_k^\sigma(x, y)},
\end{equation}
Finally we note that 
\begin{equation}
K^\sigma(x,y;t) = K_s^\sigma(x,y;t) -(-1)^{l_N} K_s^\sigma(-x,y;t),
\label{eq:image_method}
\end{equation}
as a result,
\begin{equation}\label{def:sing-extended-propagator-regularization}
\begin{aligned}
    K^\sigma(x, y; t) = &K_{\text{osc}}(x, y; t) \,
\frac{\displaystyle \sum_{k=0}^{\sigma[[[-1]]+1} Q_k^\sigma(x, y) e^{-i k t}}
{\displaystyle \sum_{k=0}^{\sigma[[[-1]]+1} Q_k^\sigma(x, y)}\\ &-(-1)^{l_N} K_{\text{osc}}(-x, y; t) \,
\frac{\displaystyle \sum_{k=0}^{\sigma[[[-1]]+1} Q_k^\sigma(-x, y) e^{-i k t}}
{\displaystyle \sum_{k=0}^{\sigma[[[-1]]+1} Q_k^\sigma(-x, y)},
\end{aligned}
\end{equation}
where $Q_k^\sigma(x, y)$ are calculated using \eqref{eq:Qk}.

\subsection{Examples of propagators \label{sec:examples}}
\begin{figure}
    \centering
    \includegraphics[width=1\linewidth]{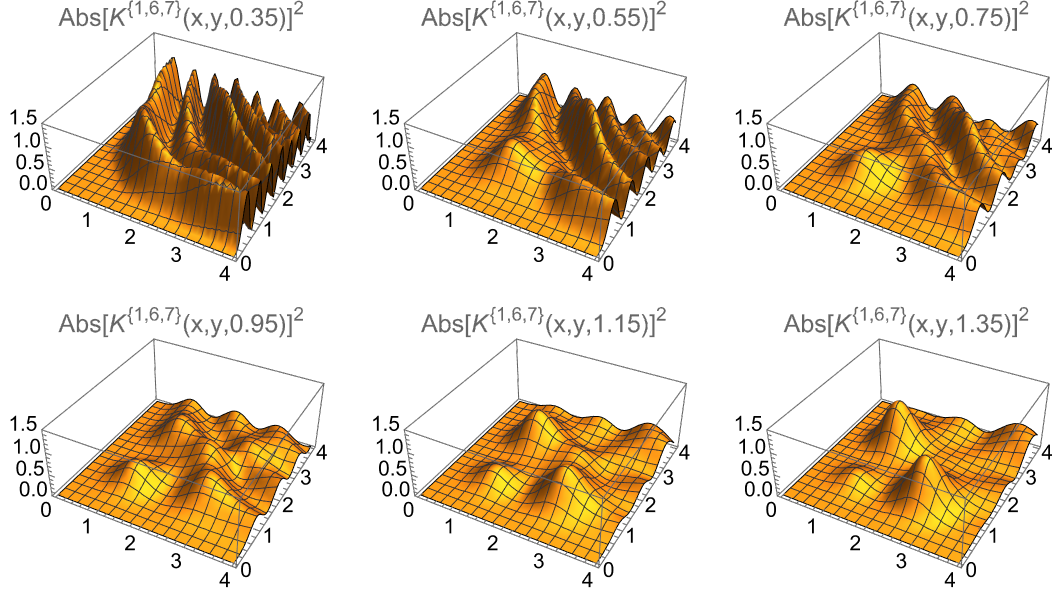}
    \caption{Transition probabilities $|K^{\{1,6,7\}}(x,y,t)|^2$ for the potential $V^{\{1,6,7\}}$}
    \label{fig:K-1-6-7}
\end{figure}

Having derived the expression for the image method in the previous section, we now proceed to apply it to specific examples of propagators. As a first illustration, we construct the propagator $K^{\{1,6,7\}}(x,y,t)$ using the image method given in equation \eqref{def:sing-extended-propagator-regularization}. Propagator $K^{\{1,6,7\}}(x,y,t)$ completely describes wave packet dynamics in the potential \eqref{eq:V-1-6-7}. In figure \ref{fig:K-1-6-7} we visualize $|K^{\{1,6,7\}}(x,y,t)|^2$ at different times to show the tunneling pattern.

\begin{figure}
    \centering
    \includegraphics[width=1\linewidth]{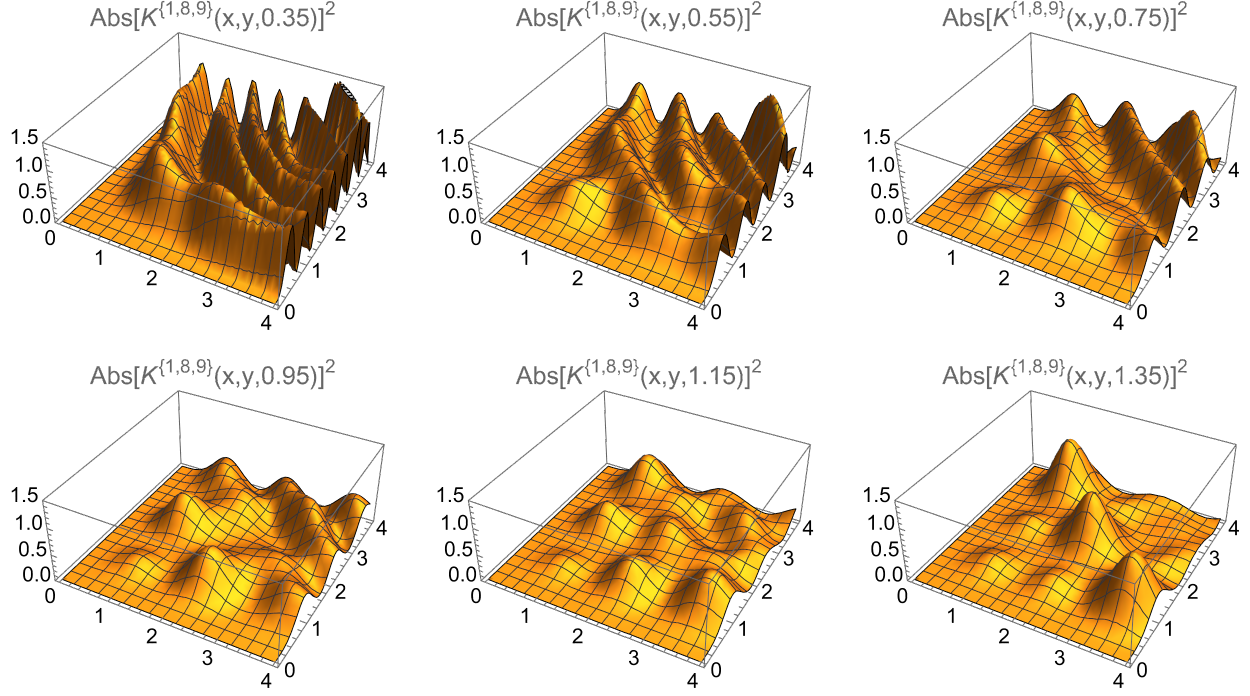}
    \caption{Transition probabilities $|K^{\{1,8,9\}}(x,y,t)|^2$ for the potential $V^{\{1,8,9\}}$}
    \label{fig:K-1-8-9}
\end{figure}

A similar analysis can be performed for the sequence $\sigma=\{1,8,9\}$. Using the same procedure, we obtain the propagator  $K^{\{1,8,9\}}(x,y,t)$. Its modulus squared is shown in Fig.~\ref{fig:K-1-8-9} at several time instants, exhibiting a distinct interference pattern.

\section{Super-partners of the Singular Oscillator and Axially Symmetric Magnetic Field configurations \label{sec:mag-field}}

In this section we provide a link between exactly solvable deformations of the singular oscillator and an axially symmetric magnetic field configurations.
This embedding provides a physical interpretation for the class of singular potentials constructed via Darboux transformations.
Within such a model, propagators that were obtained describe the radial part of the evolution of wave packets. 
Fixing $l_N$ we obtain wave packets carrying orbital momentum, so-called vortex states \cite{bliokh2012electron}.

We begin with the three-dimensional stationary Schr\"{o}dinger equation in an external magnetic field:
\begin{equation}
\frac{\left(\mathbf{p} - \frac{e}{c}\mathbf{A}\right)^2}{2m} \Psi = E \Psi\,.
\end{equation}
In cylindrical coordinate system, $\rho=\sqrt{x^2+y^2}$, $\phi$ and $z$
\begin{eqnarray}
  \mathbf{p} &=& -i\hbar \left(\partial_\rho, \frac{1}{\rho}\partial_\phi,\partial_z\right)\,,\\  
  {\rm div}\mathbf{A} &=& \frac{1}{\rho}\partial_\rho(\rho A_\rho)+ \frac{1}{\rho}\partial_\phi A_\phi +\partial_z A_z\,,\\
  \Delta &=& \frac{1}{\rho}\partial_\rho(\rho \partial_\rho)+ \frac{1}{\rho^2}\partial^2_\phi  +\partial_z^2 \,.
\end{eqnarray}

Following  \cite{bagrov2012quantum}, consider an axially symmetric magnetic field in a superposition with the Aaronov-Bohn field \footnote{The reason why we need an additional AB-field is related to the behavior of the singularity of the radial potential near the origin.}
\begin{equation}
\mathbf{A}=\mathbf{A}_m+\mathbf{A}_{AB}.    
\end{equation}
Electromagnetic potentials of AB-magnetic field along z-axis read
\begin{equation}
  A^\mu_{AB}=\left(0,0,\frac{\Phi}{2\pi\rho},0\right)\,,  
\end{equation}
The corresponding magnetic field is given by Dirac delta-function
\begin{equation}
\mathbf{B_{AB}}=(0,0,\Phi \delta(x)\delta(y))    
\end{equation}
Ratio of the AB magnetic flux to the Dirac quantum of the magnetic flux $\Phi_0=2\pi c\hbar/e$ defines the mantissa $\mu$ of the magnetic flux
\begin{equation}
    \mu=\frac{\Phi}{\Phi_0}-l_0\,,\qquad l_0= \lfloor \frac{\Phi}{\Phi_0} \rfloor \in \mathrm{Z}
\end{equation}
Thus we can write
\begin{equation}
 \frac{e}{c} A^\mu_{AB}=\left(0,0,-\hbar\frac{l_0+\mu}{\rho},0\right)\,,  
\end{equation}

The electromagnetic potentials of an axially symmetric magnetic field can be parameterized as follows
\begin{equation}
  \frac{e}{c\hbar} \mathbf{A}_m=\frac{f_2(\rho)}{\rho}\mathbf{e}_\phi
\end{equation}
\begin{equation}
 \frac{e}{c\hbar} \mathbf{B}_m=\frac{f'_2(\rho)}{\rho}\mathbf{k}\,.
\end{equation}

Using separation of variables 
\begin{equation}
    \Psi={\rm e}^{i(l-l_0)\phi}\Psi_\rho(\rho,z)\quad l\in \mathbb{Z}\,,
\end{equation}
we split off the dynamics by the angular variable
\begin{equation}
   \left[ E+\frac{\hbar^2}{2m}\left(\left(\frac{\partial^2}{\partial \rho^2}+\frac{\partial^2}{\partial z^2}\right)+
    \frac{1}{\rho}\frac{\partial}{\partial\rho}-\frac{(f_2(\rho)-l-\mu)^2}{\rho^2}\right)\right]\Psi_\rho(\rho,z)=0
\end{equation}
Further separation of radial part gives
\begin{equation}
   \frac{\hbar^2}{2m}\left[ -\frac{\partial^2}{\partial \rho^2}-
    \frac{1}{\rho}\frac{\partial}{\partial\rho}+\frac{(f_2(\rho)-l-\mu)^2}{\rho^2}\right]\psi(\rho)=E_\rho\psi(\rho)
\end{equation}
and after substitution $\psi(\rho)=u(\rho)/\sqrt{\rho}$
we arrive at the one-dimensional stationary Schr\"{o}dinger equation
\begin{equation}\label{def:radial-schr-eq}
    -u''(\rho)+V(\rho)u(\rho)=\tilde E_\rho u(\rho)
\end{equation}
\begin{equation}\label{vector-potential-to-potential}
    V(\rho)=\frac{(f_2(\rho)-l-\mu)^2-\frac{1}{4}}{\rho^2}\,,
\end{equation}
where we put $\hbar=1$, $m=1/2$.
Analytical solutions are known \cite{bagrov2012quantum} when $f_2(\rho)=\gamma\rho$ or
$f_2(\rho)=\gamma\rho^2$. The latter case, together with the choice of parameters $\gamma=1/2$, $l=0$ and $\mu=1/2$ 
leads to the oscillator potential 
\[
V_0(\rho)=\frac{\rho^2}{4}-\frac{1}{2}
\]
without centrifugal term.
In section \ref{sec:sho-def} we obtained a hierarchy of exactly solvable potentials for the radial equation 
\eqref{def:radial-schr-eq}. A natural question is what magnetic fields can produce such potentials after the separation of variables. 

We can invert \eqref{vector-potential-to-potential} to extract a vector potential from a radial potential with a centrifugal singularity at the origin
\footnote{Here we consider confining potentials with $\rho^2$ behavior at large $\rho$, therefore the centrifugal part is defined by the behavior near the origin} 
\begin{equation}\label{potential-to-vector-potential}
f_2(\rho)=  (l+\mu)+\frac{1}{2}\sqrt{1+4 \rho^2(V(\rho)+E_{reg})}  
\end{equation}
Recall, that the form of the centrifugal term 
in our hierarchy of SHO extensions 
is similar to the radial Schrodinger equation after separation of variables in spherical coordinates (because of $l_s(l_s+1)/r^2$). Centrifugal singularity in the cylindrical coordinates is different, $(l_c^2-1/4)/r^2$. To match singularities
we need $l_c=l_s+1/2$, thus both $l_c$ and $l_s$ could not be integer.
However, including Aharonov-Bohm solenoidal field and adjusting mantissa $\mu=1/2$ we can write this potential as a result of 
separation of variables of a 3D problem in the cylindrical coordinates.

From \eqref{potential-to-vector-potential} we define an axially symmetric magnetic field (without Aharonov - Bohm part)
\begin{equation}\label{potential-to-magnetic}
B_z(\rho|V(\rho),E_{reg})=\frac{2 E_{reg} + 2 V(\rho) + \rho V'(\rho)}{\sqrt{1+4 \rho^2(V(\rho)+E_{reg})}}
\end{equation}
such that the radial part of quantum dynamics with angular momentum $l$ is defined by the potential $V(\rho)$.
The parameter $E_{reg}$ should ensure that $1+4 x^2(V(\rho)+E_{reg})>0$. Since, at the level of the potential, this parameter only changes the energy reference level,
all magnetic field configurations $B_z(\rho|V(\rho),E_{reg})$ produce the same radial spectra and radial wave functions for the quantum 
states with the angular momentum $l$\footnote{Other angular momenta lead to potentials that in general are not exactly solvable and give different radial spectra and wave functions.}.
\begin{figure}
    \centering
    \includegraphics[width=0.85\linewidth]{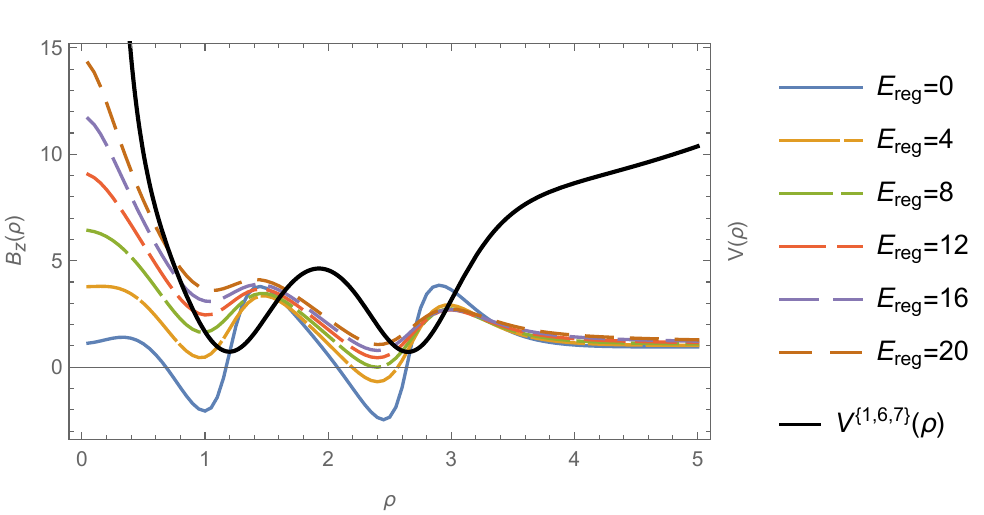}
    \caption{Radial potential $V^{1,6,7}(\rho)$ and corresponding magnetic fields $B_z(\rho,E_{reg})$ which produce isopectral $l=1$ radial states.}
    \label{fig:V-B-1-6-7}
\end{figure}

In figure \ref{fig:V-B-1-6-7} we plot radial potential $V^{1,6,7}(\rho)$ and corresponding magnetic fields $B_z(\rho,E_{reg})$. 
At large $\rho$, magnetic field $B_z(\rho,E_{reg})$ tends to a constant value and is independent from $E_{reg}$.
This is in the agreement with the equidistant spectrum of rational deformations of SHO for large values of energy.  
In the region 
of potential wells, magnetic field $B_z(\rho,E_{reg})$ oscillates following the potential (and even cylindrical layers with opposite directions of field are seen in figure \ref{fig:V-B-1-6-7} for $E_{reg}=0$).
Magnetic field at the origin 
$$
B_z(0|V^{1,6,7}(0),E_{reg})=\frac{2}{9}(5 + 3 E_{reg})
$$
can be arbitrarily large, and radial oscillations disappear for large $E_{reg}$. Still, the energy spectrum   
for $l=1$ has the ground and excited states $E_0=3+1/2$, $E_1=5+1/2$ and the equidistant part starts from $E_2=9+1/2$.
Radial ground state wave function $\psi^{\{1,6,7\}}_0(\rho)$ have two-humped structure even when $B_z(\rho|V^{1,6,7},E_{reg})$ 
approaches rapidly decreasing shape $B_z(\rho|V^{1,6,7},E_{reg}\to \infty)\to 1+\frac{\sqrt{E_{reg}}}{\rho+\rho_0}$.
It would be interesting to find radial spectra for other partial waves for a given $B_z(\rho|V^{1,6,7}(\rho),E_{reg})$ and study 
their dependence on $l$. Finally, propagators from section \ref{sec:examples} can be used to study vortex wave packets with a fixed
$l=l_N$ in magnetic fields \eqref{potential-to-magnetic}.

\section{Conclusion}

In this work, we analyzed Darboux transformations of HO into potentials with a centrifugal singularity. 
We have found, that these potentials allow more flexibility in the choice of gap sequence, beyond the well known Krein-Adler sequences. 
A possibility to organize low lying levels by choosing the gap sequence 
can be useful to model bounded systems, for instance, quarconia \cite{korolev2025multipolar} by rational extensions of SHO.
Only few levels below the charm and bottom production thresholds exist, and equidistant part of a rational extensions of SHO
can be placed above these thresholds. 

In section \ref{sec:dt-prorpagators}, we were able to calculate the quantum propagators associated with such potentials in a closed form using elementary functions.
Thus, extending results of \cite{pupasov2015propagators} to the case of singular potentials.
After a Wick-rotation, obtained propagators express transition probabilities to the diffusion processes \cite{kuznetsov2024darboux}
describing a killed Brownian motion $(0,\infty)$ with the killing
 rate $x^2/4+l_N(l_N+1)/x^2$.

Embedding the one-dimensional problem into a three-dimensional 
Schr\"{o}-dinger equation with an axially symmetric magnetic field and Aharonov--Bohm flux provide an interesting option for possible applications of SHO and analytical propagators. Equation \eqref{potential-to-magnetic} defines the axially symmetric magnetic field $B_z(\rho)$ in terms of $V^{\sigma}$. Landau levels for quantum states with $l=l_N$ in such a field are defined by 
the gap sequence $\sigma$. 
Moreover, axially symmetric magnetic-field configurations obtained in the section \ref{sec:mag-field} may be applied to study
propagation of vortex packets \cite{sizykh2024nonstationary,sizykh2024transmission} by using corresponding analytical propagators. Usually, such vortex states are modeled by Laguerre-Gaussian packets.
Our rationally extended SHO gives rise to the {\it exceptional-Laguerre-Gaussian} packets, following the terminology of the theory of exceptional orthogonal polynomials.

Note, that \eqref{potential-to-magnetic} is not restricted to analytically solvable potentials generated by Darboux transformations. 
It can be applied to a rather general potential, thus producing iso-spectral magnetic field configurations, parametrized by $E_{reg}$ for the quantum states with a fixed $l$. 

In addition to the analytical treatment of the singular harmonic oscillator potentials and their corresponding propagators, we provide a set of numerical simulations illustrating the eigenstate structure and the interference patterns produced by their rational extensions. These visualizations support the internal consistency of our closed-form expressions and reveal nontrivial dynamical behaviors induced by different gap sequences. To facilitate reproducibility, the supplementary repository \cite{pupasovmaksimov_prop_reho_2025} includes a Wolfram Mathematica notebook, together with sample plots, that computes the propagators for the rationally extended singular harmonic oscillator.

\section*{Acknowledgements}
We thank our families and friends for their support. AMPM thanks 
the Mathematics Department at UFJF and the Theoretical and Mathematical physics Laboratory at TSU for the opportunity of scientific visit.
We are grateful to professor P.O. Kazinski for helpful discussions.
This research did not receive any specific grant from funding agencies in the public, commercial, or not-for-profit sectors.



\bibliographystyle{unsrtnat} 
\bibliography{example}






\end{document}